\documentclass{PoS}

\title{Status of ProtoDUNE Dual Phase}

\ShortTitle{Status of ProtoDUNE Dual Phase}

\author{\speaker{C. Cuesta} on behalf of DUNE collaboration \\
        Centro de Investigaciones Energ$´{e}$ticas, Medioambientales y Tecnol$´{o}$gicas,
 CIEMAT, \\ 28040, Madrid, Spain\\
        E-mail: \email{clara.cuesta@ciemat.es}}

\abstract{The Deep Underground Neutrino Experiment (DUNE) is a dual-site experiment for long baseline neutrino oscillation studies, and for neutrino astrophysics and nucleon decay searches. DUNE will comprise four 10 kton fiducial liquid argon time-projection-chamber (LAr TPC) modules placed at the Sanford Underground Research Facility (South Dakota, USA). One of these modules will profit form the dual phase (DP) technology where the charge is extracted, amplified, and detected in gaseous argon above the liquid surface allowing a fine readout pitch, a low energy threshold, and good pattern reconstruction of the events. To gain experience in building and operating such a large-scale DP LAr detector, a prototype is currently being assembled at the CERN Neutrino Platform. The ProtoDUNE-DP detector consists of a 6x6x6 m$^3$ LAr TPC and commissioning started in Summer 2019.}

\FullConference{
European Physical Society Conference on High Energy Physics - EPS-HEP2019 -\\
			10-17 July, 2019\\
			Ghent, Belgium}

\begin{document}

\section{The Deep Underground Neutrino Experiment (DUNE) project}
The DUNE experiment aims to address key questions in neutrino physics and astroparticle physics~\cite{duneIDRv1}. It includes precision measurements of the parameters that govern neutrino oscillations with the goal of measuring the CP violating phase and the neutrino mass hierarchy with a muon neutrino beam produced at Fermilab. The physics programme also addresses non-beam physics as nucleon decay searches and the detection and measurement of the electron neutrino flux from a core-collapse supernova within our galaxy. DUNE will consist of a near detector placed at Fermilab close to the production point of the muon neutrino beam of the Long-Baseline Neutrino Facility (LBNF), and four 10\,kt fiducial mass LAr TPCs as far detector in the Sanford Underground Research Facility (SURF) at 4300\,m.w.e. depth at 1300\,km from Fermilab~\cite{duneIDRv2,duneIDRv3}. 

In order to gain experience in building and operating such large-scale LAr detectors, an R\&D programme is currently underway at the CERN Neutrino Platform.  Such programme operates two prototypes with the specific aim of testing the design, assembly, and installation procedures, the detector operations, as well as data acquisition, storage, processing, and analysis.  The two prototypes employ LAr TPCs as detection technology.  One prototype only uses LAr, called ProtoDUNE Single-Phase, and the other uses argon in both its gaseous and liquid state, thus the name ProtoDUNE Dual-Phase.  Both detectors have similar sizes. In particular, ProtoDUNE-DP~\cite{wa105} has an active volume of 6$\times$6$\times$6~m$^3$ (300\,tonne).

\section{Dual-Phase (DP) LAr TPC}

Charged particles that traverse the active volume of the LAr TPC ionize the medium, while also producing scintillation light. The ionization electrons drift along an electric field towards a segmented anode where they deposit their charge, and photon detectors pick up the scintillation light.  The DP  design allows for a single, fully homogeneous LAr volume with a much longer drift length. The volume is surrounded by a field cage on the sides and a cathode at the bottom, which together define the drift field. In a DP LAr TPC the ionization charges are amplified in the ultra-pure cold argon vapor layer above the liquid to obtain low-energy detection thresholds with high signal-to-noise ratio over drift distances exceeding 10~m. 
 
\begin{figure}[ht]
\centering
 \includegraphics[width=0.48\textwidth]{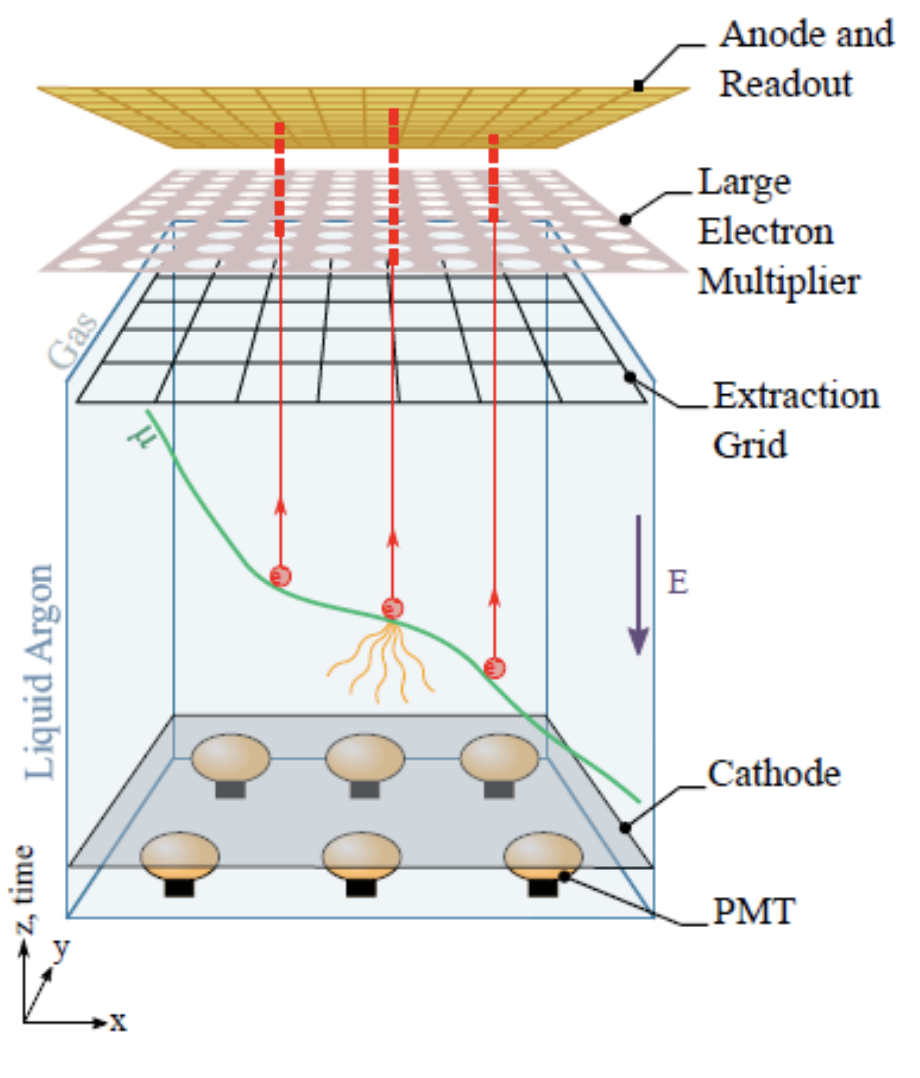}
\caption{Principle of the DP readout}
\label{DP}
\end{figure}

The schematic drawing on Figure~\ref{DP} summarizes the elements of the DP LAr TPC.  The key differentiating concept of the DP design is the amplification of the ionization signal in an avalanche process. Electrons drift vertically towards an extraction grid just below the liquid-vapor interface.  After reaching the grid, an electric field stronger than the drift field extracts the electrons from the liquid into the gas phase. Once in the gas, electrons encounter micro-pattern gas detectors with high-field regions, called large electron multipliers (LEMs).  The LEMs amplify the electrons in avalanches that occur in these high-field regions and a secondary electroluminescence light is produced. The amplified charge is then collected and recorded on a 2D anode and the light by the photon detection system.  The extraction grid, LEM and anode are assembled into three-layered sandwiches called charge-readout planes (CRPs). The precision tracking and calorimetry offered by the DP technology provides excellent capabilities for identifying interactions of interest while mitigating sources of background. 


\section{The 3\,$\times$\,1\,$\times$\,1\,m$^3$ DP LAr TPC}

\begin{figure}[ht]
\centering
 \includegraphics[width=0.8\textwidth]{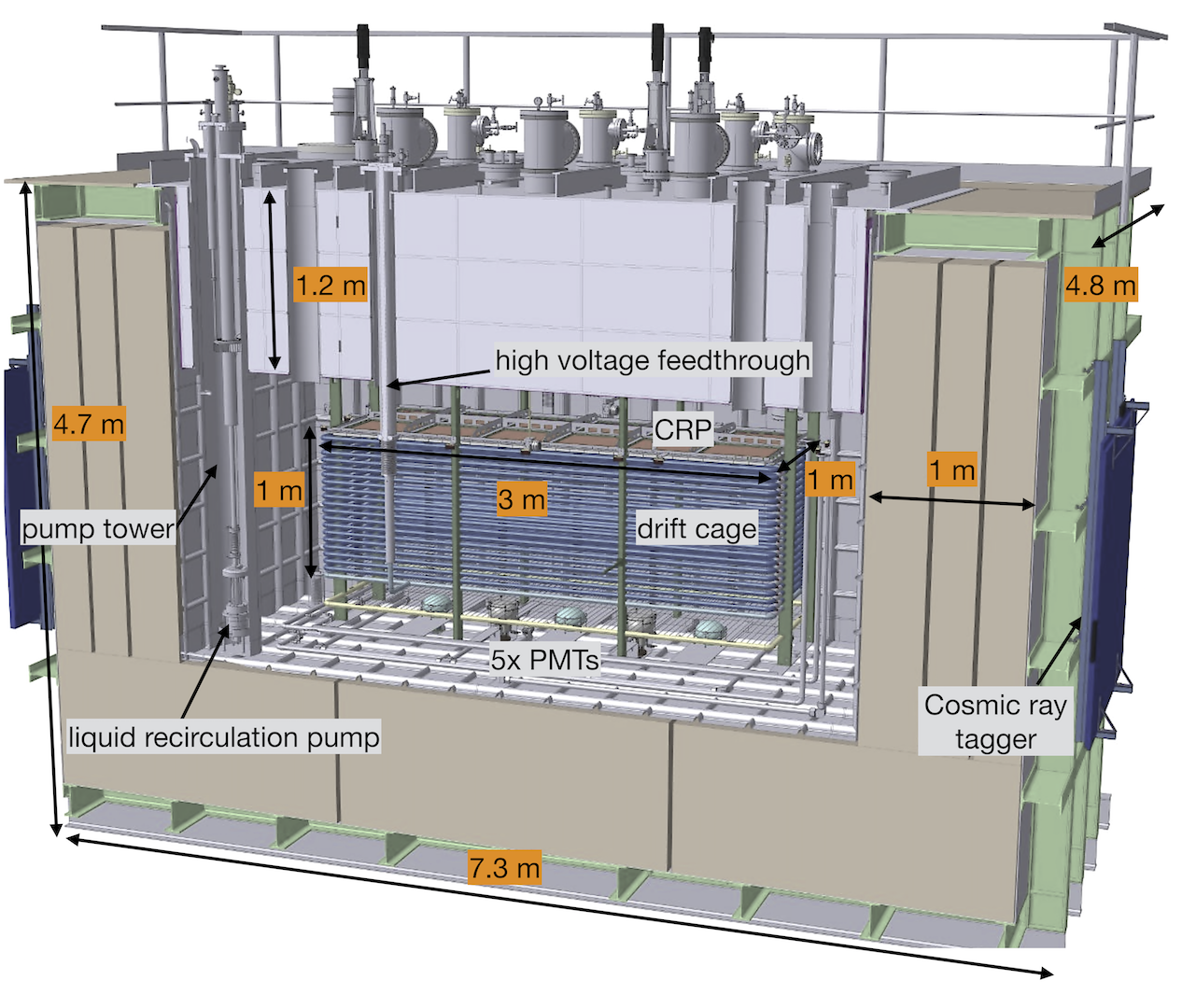}
\caption{Drawing of the 3\,$ \times$\,1\,$ \times$\,1\,m$^3$ dual-phase LAr TPC in the cryostat.}
\label{311}
\end{figure}

To validate the concept of a non-evacuated industrial cryostat, test several key sub-systems for ProtoDUNE-DP, and demonstrate for the first time the capabilities of the dual-phase LAr TPC technology on a tonne scale, a demonstrator with an active volume of 3$\times$1$\times$1~m$^3$ (4.2~tonne) was built in 2016 and operated in 2017 at CERN~\cite{311}. The 3$\times$1$\times$1~m$^3$ TPC is shown in Fig.~\ref{311}. The detector is suspended under a 1.2~m  thick insulating lid called top-cap where the necessary feedthroughs are hosted. The electron extraction, amplification, and collection are performed inside a 3$\times$1~m$^2$ CRP.  The CRP is electrically and mechanically independent from the drift cage and can be remotely adjusted to the liquid level. The detection of the scintillation light is done at the bottom of the detector by an array of five cryogenic photomultipliers (PMTs).  

The newly employed membrane cryostat provided a stable cryogenic environment with a flat liquid surface allowing for charge extraction over an area of 3\,m$^2$. An excellent LAr purity was also achieved with a corresponding electron lifetime of around 4 ms. While data collection over brief time periods was possible, the high voltage instabilities in the operation of the extraction grid prevented a proper study of the detector long-term stability and performance. The maximum stable amplification field that could be reached in the LEMs was also lower than envisioned for large DP LAr TPCs. For ProtoDUNE-DP, the improvements on the designs of the extraction grid mounting and its HV connection and LEMs address these issues. A first look at the data collected at an effective gain of around 3 before complete charging up of the LEMs nevertheless demonstrated the high quality of the DP TPC imaging. We also demonstrated the charge readout with two collection planes with strips of up to 3\,m length and almost equalized charge collection. We were able to detect the prompt and electroluminescence scintillation and observed a correlation between electroluminescence in the gas and collected charge.

\section{ProtoDUNE Dual Phase}

The ProtoDUNE-DP experiment at CERN has the main goal to validate both the DP design for the DUNE far detector~\cite{duneIDRv3} and provide insight on the detector response with cosmic rays. ProtoDUNE-DP has an active mass of 300\,tonne and a total drift distance of 6 m (one-half of the total drift at DUNE). In the following subsections the main components of ProtoDUNE-DP are described.

\subsection{Charge-Readout Planes}

\begin{figure}[ht]
\centering
 \includegraphics[width=0.9\textwidth]{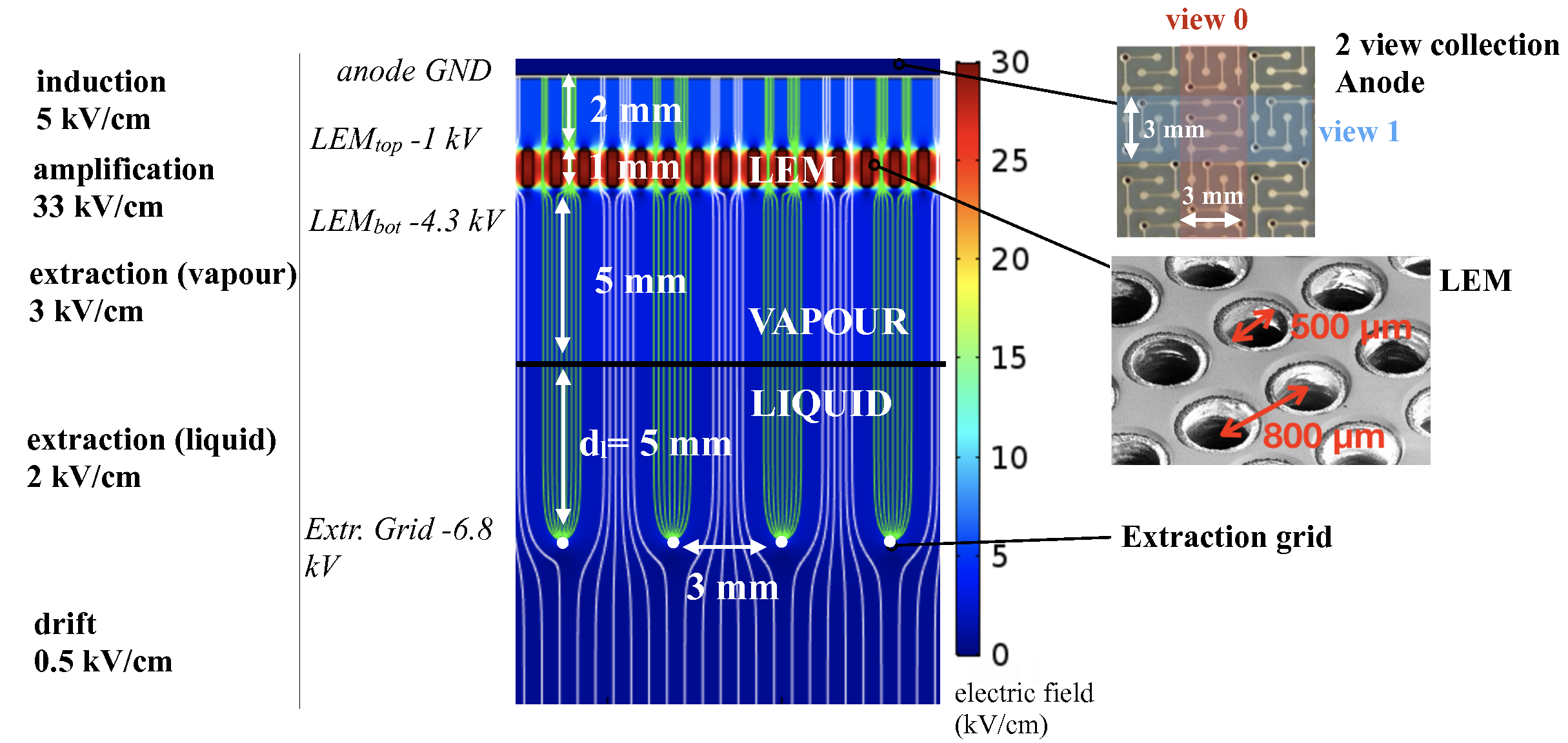}
\caption{Illustration of the electric fields in the amplification region of ProtoDUNE-DP. The simulated field
lines in dark blue indicate the paths followed by the drifting charges.}
\label{lem}
\end{figure}

An extraction efficiency of 100\% of the electrons from the liquid to the gas phase is achieved with an electric field of the order of 2 kV/cm across the liquid-gas interface, applied between an extraction grid submersed in the liquid and charge amplification devices situated in the ultra-pure argon gas. The amplification devices, called LEMs, are horizontally oriented 1 mm-thick printed circuit boards with electrodes on the top and bottom surfaces. They are drilled through with many holes that collectively form a micro-pattern structure; when a 3\,kV potential difference is applied across the electrodes the ionization electrons are amplified by avalanches (Townsend multiplication) occurring in the pure argon gas in this micro-pattern structure due to the high electric field (30\,kV/cm).  The charge is collected in a finely segmented 2D (x and y) readout anode plane at the top of the gas volume and fed to the frontend electronics. A diagram of the electric fields in the amplification region of ProtoDUNE-DP is shown in Figure~\ref{lem}.

The collection, amplification and readout components are combined in the CRPs, an array of independent (layered) modules. A CRP is composed of 36 0.5\,$\times$\,0.5 m$^2$ units, each of which is composed of a LEM-anode sandwich. The CRP structure also integrates the submersed extraction grid, which is an array of x and y oriented stainless steel wires, 0.1\,mm in diameter, with 3.125\,mm pitch. In ProtoDUNE-DP, 2 fully instrumented CRPs, and 2 CRPs without LEMs are installed.

\subsection{Readout electronics}

The electrical signals from the collected charges are passed to the outside of the tank via a signal feedthrough chimney. The cryogenic front-end electronics cards, housed at the bottom of the chimney are based on analog cryogenic preamplifiers implemented in CMOS ASIC circuits for high integration and large-scale affordable production. Within the chimney, the cards are actively cooled to a temperature of about 110 K and isolated with respect to the LAr vessel by the cold flange feedthrough.  

The digital electronics is located on the roof the cryostat at room temperature. This makes it possible to use Micro Telecommunications Computing Architecture ($\mu$TCA) Digitization cards in the advanced mezzanine card (AMC) format to read 64 channels per card. Each AMC card can digitize all 64 channels at 2.5 MHz and compress and transmit this continuous data stream, without zero skipping, over a network link operating at 10 Gbit/s.

\subsection{Cathode, Field Cage and HV System}

The drift field (0.5 kV/cm) inside the fully active LAr volume is produced by applying HV to the cathode plane at the bottom of the cryostat and is kept uniform by the field cage, a stack of equally spaced field-shaping electrodes.  The cathode structure, constructed of a reinforced frame to guarantee its planarity, is suspended from the field cage and hangs near the bottom of the cryostat. It is a segmented structure of tubes of different sizes arranged in a grid to minimize weight and allow scintillation light to pass through and detected by the PMTs mounted at the bottom of the tank.

\subsection{Photon Detection System}

The photon detection system of ProtoDUNE-DP~\cite{protoDUNElight}  is formed by 36 8-inch cryogenic PMTs (R5912-02MOD from Hamamatsu) placed below the cathode grid. The ProtoDUNE-DP PMTs were validated and characterized~\cite{protoDUNEPMTs} . As wavelength shifter, a sheet of polyethylene naphthalate (PEN) is placed on top of 30 PMTs and the other 6 PMTs have  tetraphenyl butadiene (TPB) directly coated on them. There is a LED-based fiber calibration system to monitor the PMT performance and obtain an equalized PMT response~\cite{protoDUNELCS}.

The scintillation light signal is used as trigger for non-beam events, to determine precisely the event time, needed for a full 3D reconstruction of non-beam events, and for cosmic background rejection. There might be also a possibility to perform particle identification exploiting the detected light signals.

\subsection{Cryogenic instrumentation and slow control}

The cryogenic instrumentation and slow controls system provides comprehensive monitoring for all detector module components as well as for the LAr quality and behavior, both being crucial to guarantee high-quality of the data.  The cryogenic instrumentation includes a set of devices to monitor the quality and behavior of the LAr volume in the cryostat interior, ensuring the correct functioning of the full cryogenics system and the suitability of the LAr for good quality physics data. These devices  include purity monitors, temperature monitors, LAr level monitors, and cryogenic cameras with their associated light emitting system.

\section{Prospects}

\begin{figure}[ht]
\centering
 \includegraphics[width=0.44\textwidth]{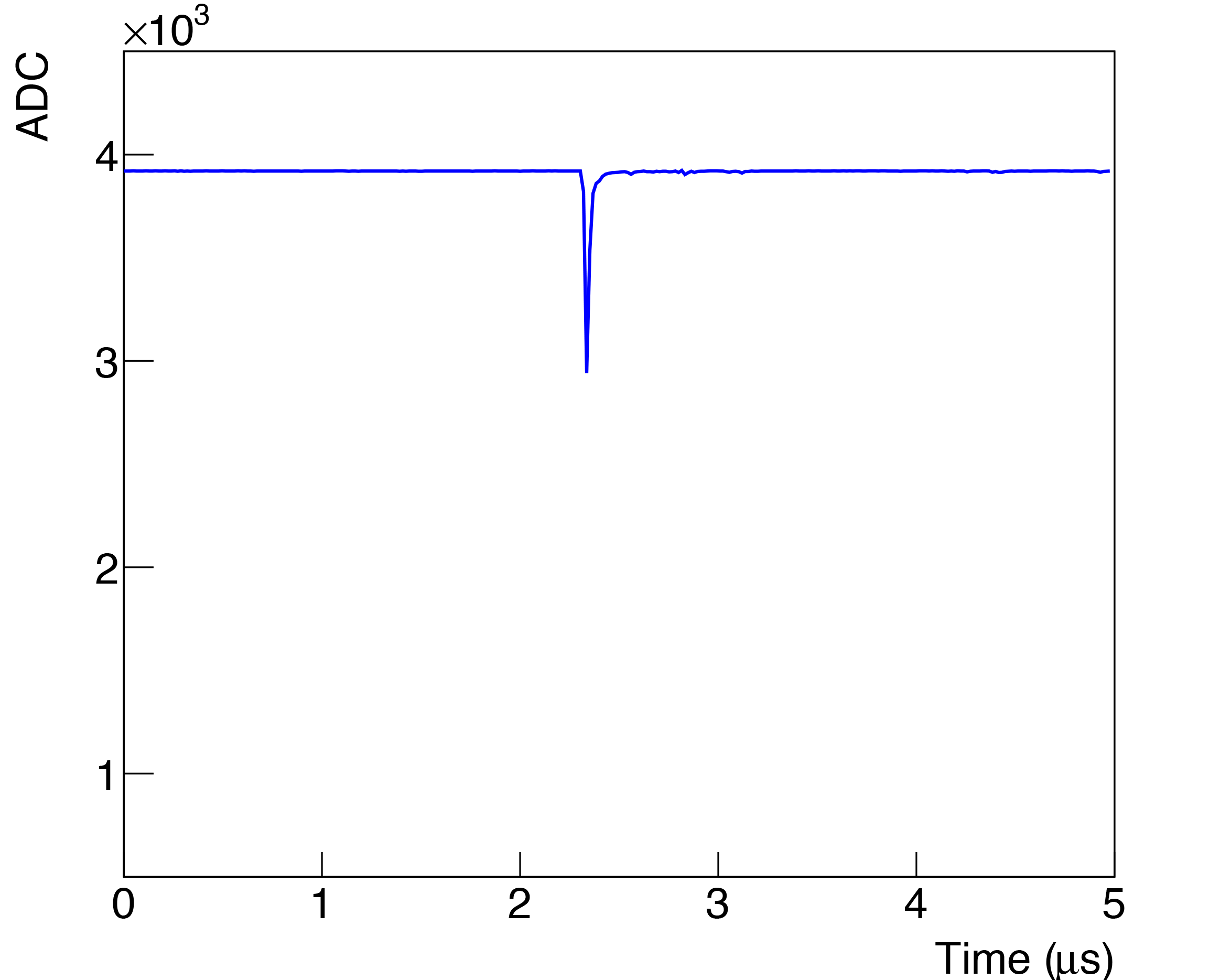}
 \includegraphics[width=0.44\textwidth]{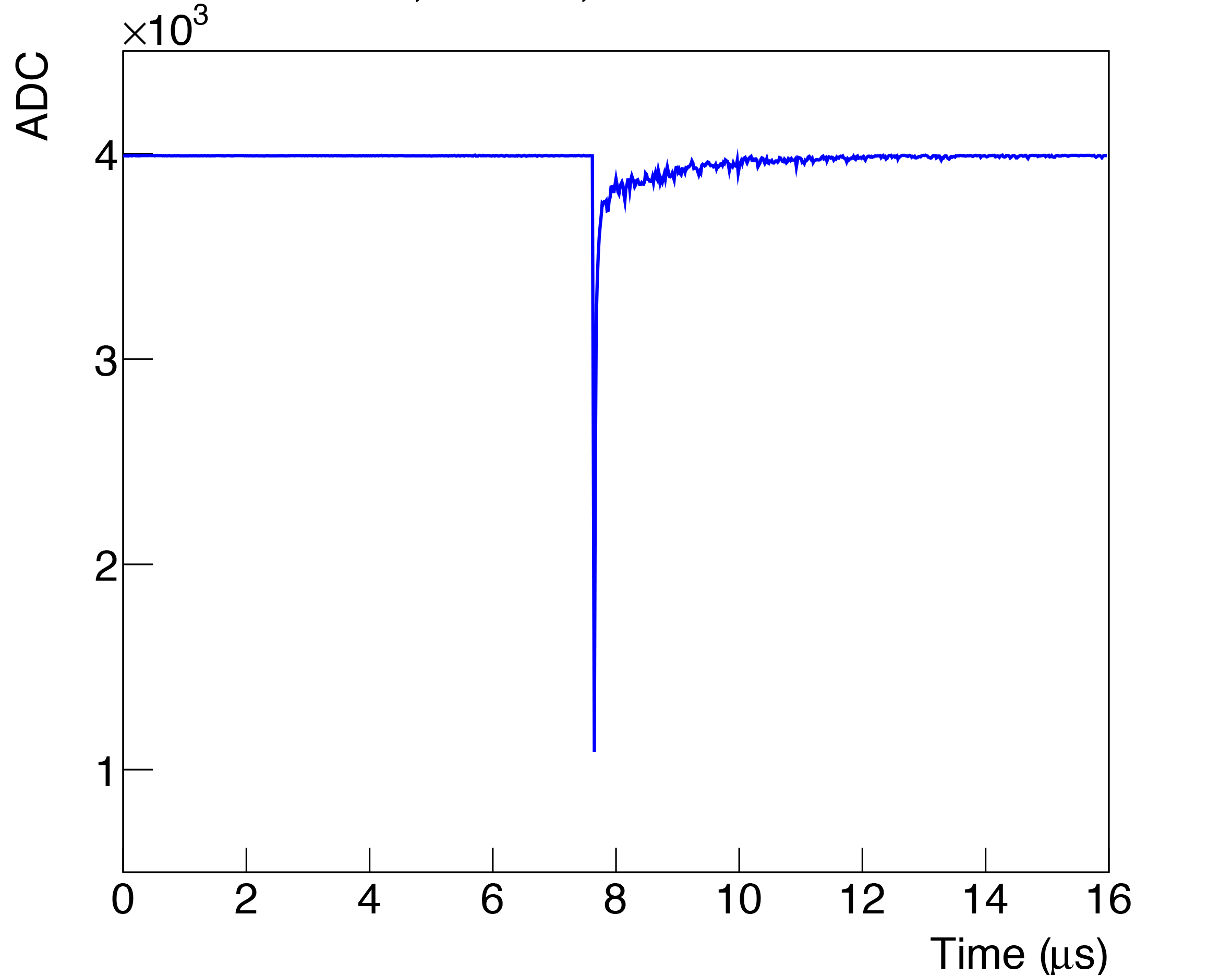}
\caption{Examples of scintillation light events in one PMT of ProtoDUNE-DP when the detector was filled with gaseous argon (left) and with LAr (right).}
\label{WF}
\end{figure}

During Summer 2019, the cryostat was closed, purged with gas argon, and filled with LAr. Light data taking started in gas and continued in LAr, see event examples in Figure~\ref{WF}. At the moment, ProtoDUNE-DP is being commissioned, and  will take charge and light data with cosmic rays during several months. An upgrade for a phase II  to instrument all 4 CRPs among other improvements is envisioned. The results will be key for the construction of the DUNE far detector DP module.

\acknowledgments
This project has received funding from the European Unions Horizon~2020 Research and Innovation programme under Grant Agreement no.~654168 and from the Spanish Ministerio de Econom$´{i}$a y Competitividad (SEIDI-MINECO) under Grants no. FPA2016-77347-C2-1-P and MdM-2015-0509.

\bibliographystyle{JHEP}
\bibliography{ProtoDUNE_DP_EPS}



\end{document}